\documentclass[a4paper,11pt]{article}
\usepackage{pos}
\usepackage{slashed}
\usepackage{cancel}

\newcommand{\ii}{{\overline{\imath}}}

\title{On the double counting subtraction at NLO${}^\star$ of the high-energy factorization approach}

\author*[a]{A.~Chernyshev}
\author[a]{V.~Saleev}

\affiliation[a]{Bogoliubov Laboratory of Theoretical Physics, Joint Institute for Nuclear Research, Dubna, Russia}

\emailAdd{chernyshev@theor.jinr.ru}
\emailAdd{saleev.vladimir@gmail.com}

\FullConference{QCD at the Extremes, 1-5 September, 2025}

\abstract{
We suggest improvements for double counting subtraction scheme, which is needed for the consistent treatment of real corrections in the high-energy limit, and apply it to the single photon production at the NLO${}^\star$ approximation of the high-energy factorization approach.
The presented improvements allow us to avoid the oversubtraction problem.
}

\begin{document}
\maketitle

\section{Introduction}

The advantages in usage of high-energy factorization (HEF) approach~\cite{HEF} for description of various hard processes have been demonstrated, for example, in recent Refs.~\cite{HEF2}, so the motivation for extending this approach beyond the leading order (LO) approximation in $\alpha_s$ is clear; a short time ago, the scheme of the next-to-leading order (NLO) calculations has been developed~\cite{HN25}.
In this context, the non-trivial problem of double counting between real NLO corrections (NLO${}^\star$) and unintegrated PDFs (UPDFs), which are responsible for the resummation of initial state radiations in the HEF, appears and investigation of the corresponding subtraction scheme is necessary~\cite{DC2}.
A similar procedure for localization of NLO${}^\star$ corrections to impact factors (IF) to avoid double counting with Green's function has also been developed~\cite{DC} within the BFKL formalism~\cite{BFKL}.
However, it turns out that in some cases this procedure can lead to oversubtraction, especially at low rapidity differences.
In this paper, we provide a simple improvement of this procedure, thus eliminating oversubtraction, and apply it to the production of single photon~\cite{PRA2}, which is a convenient process for studying the subtraction terms due to the infrared (IR) finiteness of one of the NLO${}^\star$ corrections.

\section{HEF}

For instance, we consider hadroproduction of the single photon $p \, (P_1) + p \, (P_2) \to \gamma \, (q_3) + X$\footnote{Light-cone vectors can be introduced such that $n_\mp = \left( 2 / \sqrt{S} \right) P_{1, 2}$, so $q = q_L + q_T$, where $q_L = \left( q^+ n_- + q^- n_+ \right) / 2$ with $q^\pm = (q, n_\pm)$, rapidity $y(q) = (1 / 2) \ln\left( q^+ / q^- \right)$.}, the cross section for this process within the HEF reads:
\begin{equation}
\sigma =
\int \frac{d x_1 }{x_1} \frac{d x_2}{x_2} \frac{d^2 {\bf q}_{T 1}}{\pi} \frac{d^2 {\bf q}_{T 2}}{\pi}\
\Phi_{i_1}(x_1, {\bf q}_{T 1}^2, \mu^2) \,
\hat\sigma_{i_1 i_2}(x_{1, 2}, {\bf q}_{T 1, 2}, \bar\alpha_s(\mu_R^2)) \,
\Phi_{i_2}(x_2, {\bf q}_{T 2}^2, \mu^2);
\label{eq:HEF}
\end{equation}
adopting the parton Reggeization approach~\cite{PRA}, hard scattering coefficient (HSC) $\hat\sigma_{i_1 i_2}$ describing subprocess $i_1 \, (q_1) + i_2 \, (q_2) \to \gamma \, (q_3)$ is calculated with Reggeized partons $i_{1, 2}$, i.e., Reggeized gluons $R$ and quarks $Q$; $q_1 = \left( q_1^+ / 2 \right) n_- + q_{T 1}$ and $q_2 = \left( q_2^- / 2 \right) n_+ + q_{T 2}$, where $q_1^+ = x_1 P^+$ and $q_2^- = x_2 P^-$.
In Eq.~(\ref{eq:HEF}), the UPDFs are defined as follows:
\begin{equation}
\Phi_i(x, {\bf q}_T^2, \mu^2) = \sum_{\ii}
\int\limits_0^1 \frac{d z}{z} \
\theta\left( \Delta(|{\bf q}_T|, \mu) - z \right) \,
C_{i \ii}(z, {\bf q}_T^2, \mu^2) \,
\tilde f_\ii\left( \frac{x}{z}, \mu^2 \right),
\label{eq:UPDF}
\end{equation}
where $z$ is a fraction of the large light-cone component carried through resummation in $C_{i \ii}$, $\tilde f_i\left(x, \mu^2 \right) = x f_i\left(x, \mu^2 \right)$, and $\mu$ is a typical hard scale; in Eq.~(\ref{eq:UPDF}), the cutoff $\Delta(|{\bf q}_T|, \mu) = \mu / \left( \mu + |{\bf q}_T| \right)$ ensures rapidity ordering between partons emitted in the UPDF and HSC in Eq.~(\ref{eq:HEF}).

\begin{figure}
\centering
\includegraphics[scale=1.]{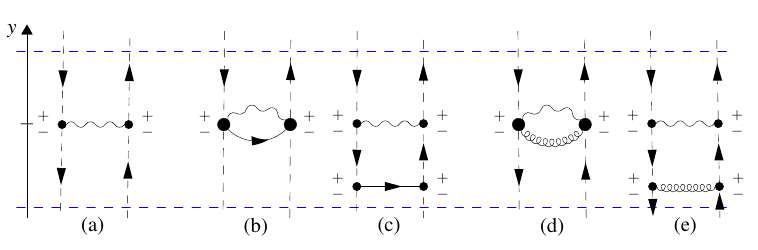}
\caption{LO, NLO${}^\star$ and corresponding subtraction terms plotted on the rapidity axis. Blue lines denotes boundaries of forward / backward regions.}
\label{fig:1}
\end{figure}

\section{LO, NLO${}^\star$, and double counting subtraction}

The HSCs for the LO (Fig.~\ref{fig:1}~(a)) and NLO${}^\star$ (Fig.~\ref{fig:1}~(b),~(d)) subprocesses
\begin{eqnarray}
Q \, (q_1) + \bar Q \, (q_2) &\to &\gamma \, (q_3),
\label{eq:LO1} \\
Q \, (q_1) + R \, (q_2) & \to & \gamma \, (q_3) + q \, (q_4),
\label{eq:NLO1} \\
Q \, (q_1) + \bar Q \, (q_2) & \to & \gamma \, (q_3) + g \, (q_4),
\label{eq:NLO2}
\end{eqnarray}
already were calculated in Refs.~\cite{PRA2}.
The subprocesses~(\ref{eq:NLO1}),~(\ref{eq:NLO2}) have double counting with the subprocess~(\ref{eq:LO1}) when integrating over the entire $d y_4 = d \hat z / (1 - \hat z)$, where $y_4 = \ln\left( |{\bf q}_{T 4}| / \left( (1 - \hat z) \, q_2^- \right) \right)$ and $\hat z = q_3^- / q_2^-$, such that the additional parton $i = q, g$ goes deeply forward / backward in rapidity space, so it should be included in the UPDF~(\ref{eq:UPDF}).
This contributions can be subtracted as it shown in Fig.~\ref{fig:1}~(c),~(e) (the permutations $+ \leftrightarrow -$ are assumed), the gauge-invariant amplitudes reads:
\begin{eqnarray}
A^\mu_a(\text{Fig.~\ref{fig:1}~(c)}) & = &
e g T_a \, \Theta(\hat z)
\left( \frac{q_2^-}{2 |{\bf q}_{T 2}|} \right)
\bar u (q_4) \, \gamma_+(q, - q_4) \, \frac{1}{\slashed{q}} \, \hat P^+ \, \Gamma^\mu(q_1, -q) \, u(q_{L 1}),
\label{eq:A1} \\
A^{\mu \nu}_a(\text{Fig.~\ref{fig:1}~(e)}) & = &
e g T_a \, \Theta(\hat z) \,
\bar u (q_{L 2}) \, \Gamma^\nu(q, q_2) \, \frac{1}{\slashed{q}} \, \hat P^+ \, \Gamma^\mu(q_1, -q) \, u(q_{L 1}),
\label{eq:A2}
\end{eqnarray}
where $q = q_1 - q_3$, propagator of the Reggeized quark is $\theta(Y) \left( i / \slashed{q} \right) \hat P^+$, $\hat P^+ = (1 / 4) \, \slashed{n}_- \slashed{n}_+$, and
\begin{eqnarray*}
\gamma_+(q, - q_4) & = &
\slashed{n}_+ + \slashed{q} \, \frac{2}{(q - q_4)^-}, \quad
\\
\Gamma^\mu(q_1, -q) & = &
\gamma^\mu - \slashed{q}_1 \, \frac{n_-^\mu}{(q_1 - q)^-} + \slashed{q} \, \frac{n_+^\mu}{(q_1 - q)^+};
\end{eqnarray*}
in Eqs.~(\ref{eq:A1}),~(\ref{eq:A2}) cutoff function
\begin{equation}
\Theta(\hat z) =
\theta\left( \frac{|{\bf q}_{T 3}|}{|{\bf q}_{T 4}|} - \hat z \right)
\theta\left( \frac{|{\bf q}_{T 2}| |{\bf q}_{T 3}|}{\Delta(|{\bf q}_{T 2}|, \mu) \, {\bf q}_{T 4}^2} - \hat z \right)
\label{eq:Theta}
\end{equation}
has the following origin.
The propagator contain $\theta$--function of
\begin{equation*}
Y = y(q_3) - y(q_4) =
\ln\left( \frac{1 - \hat z}{\hat z} \frac{|{\bf q}_{T 3}|}{|{\bf q}_{T 4}|} \right),
\end{equation*}
which divide photon and parton $i$ by rapidities, this generates the first factor in Eq.~(\ref{eq:Theta}).
However, the real correction still can be far from the UPDF in the rapidity space due to the Reggeized parton $i_2$, so it seems natural at least to require $Y > Y_2$ also, where rapidity difference between parton $i$ and last emitted parton in $C_{i_2 \ii_2}$~(\ref{eq:UPDF}) reads:
\begin{equation*}
Y_2 =
\ln\left( \frac{z_2 / (1 - z_2)}{1 - \hat z} \frac{|{\bf q}_{T 4}|}{|{\bf q}_{T 2}|} \right),
\end{equation*}
here $z_2$ is the same as $z$ in Eq.~(\ref{eq:UPDF}) for UPDF $\Phi_{i_2}$ in Eq.~(\ref{eq:HEF}).
Probably, the most accurate way is to amend the HEF factorization~(\ref{eq:HEF}) for subtraction terms by keeping the $z_2$--dependence in HSC through $\theta(Y - Y_2)$.
Another way is to keep Eq.~(\ref{eq:HEF}) by estimating $z_2 \sim \Delta(|{\bf q}_{T 2}|, \mu)$, see Eq.~(\ref{eq:UPDF}), which leads to the second factor in Eq.~(\ref{eq:Theta}).

\begin{figure}
\centering
\includegraphics[scale=.35]{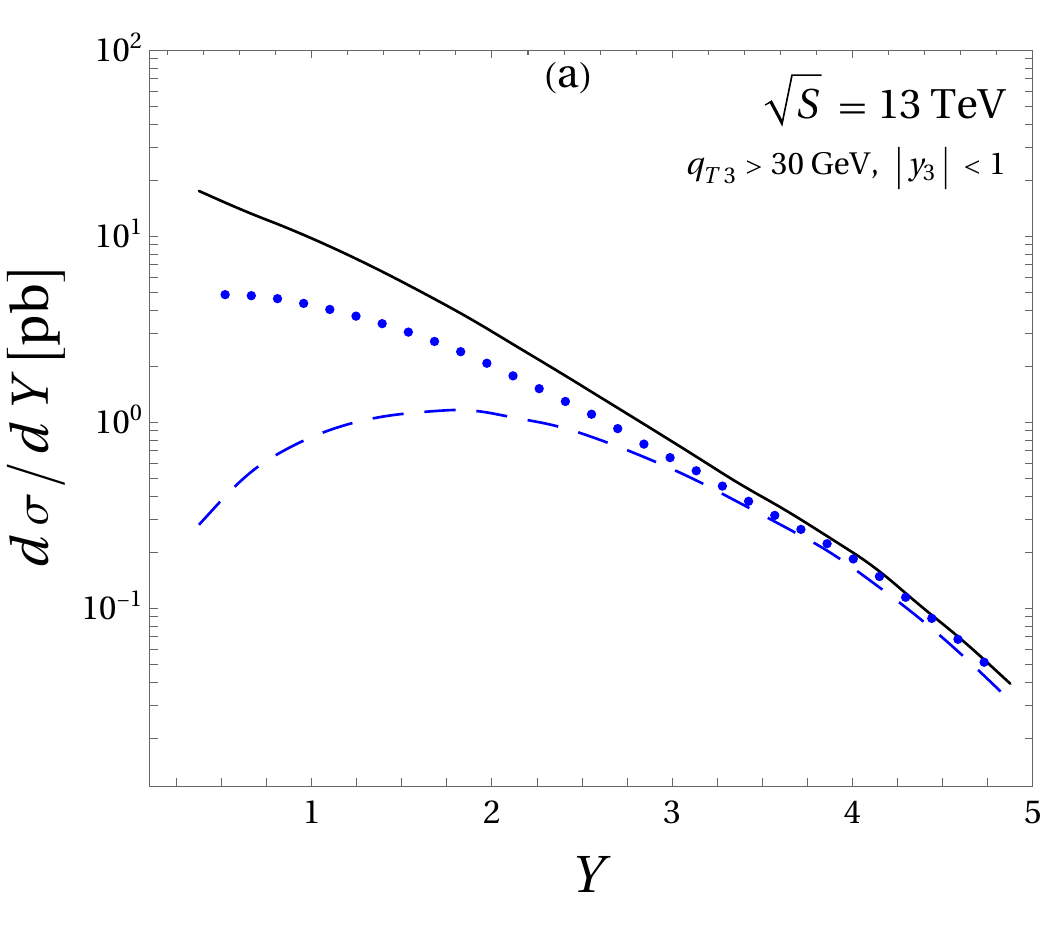}
\qquad
\includegraphics[scale=.35]{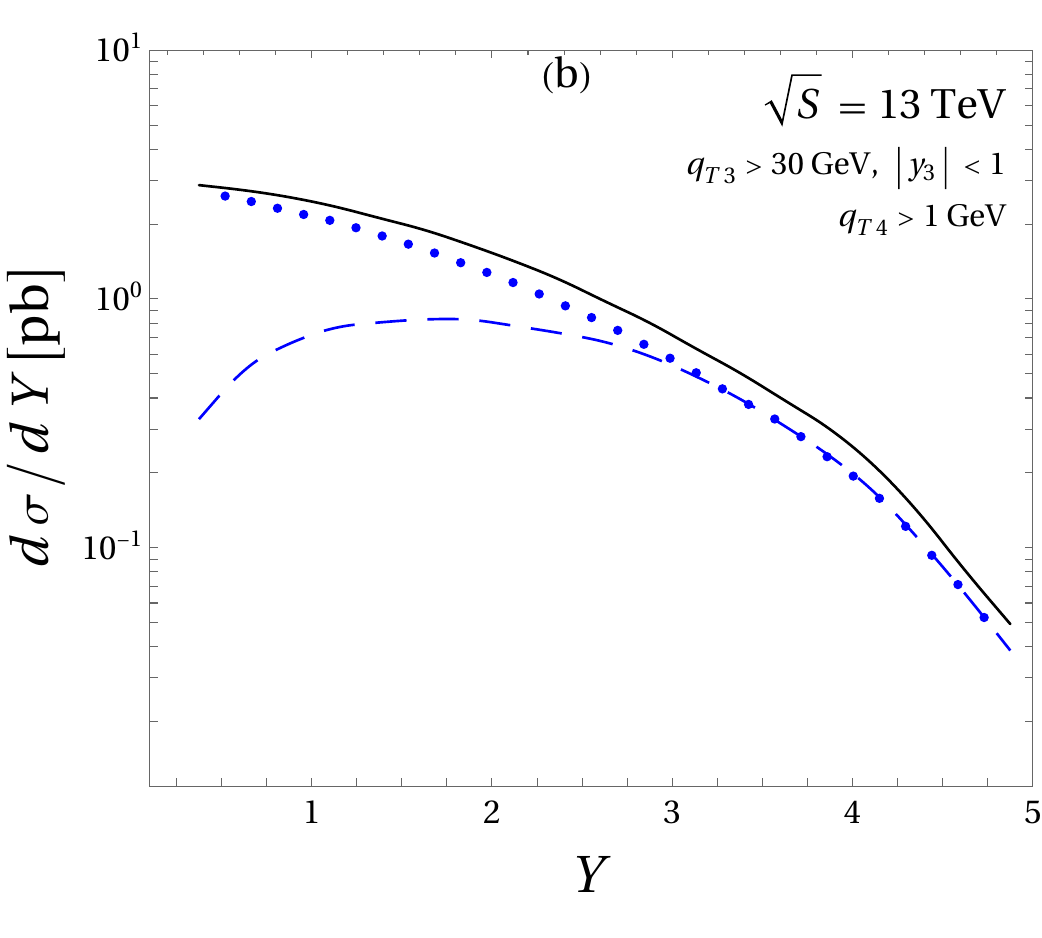}
\caption{Cross section as function of $Y$ (a) for~(\ref{eq:NLO1}) and (b) for~(\ref{eq:NLO2}).
Different contributions are shown: NLO${}^\star$ (black lines) and subtraction terms without / with second factor in Eq.~(\ref{eq:Theta}) (dotted / dashed blue lines).}
\label{fig:2}
\end{figure}

The HSCs corresponding to Eqs.~(\ref{eq:A1}),~(\ref{eq:A2}) can be written as follows:
\begin{eqnarray}
\frac{d \hat\sigma\left( Q \, (q_1) + i_2 \, (q_2) \to \gamma \, (q_3) + i \, (q_4) \right)}{d {\bf q}_{T 3}^2 d y_3 d {\bf q}_{T 4}^2 d \Delta\phi_{3 4}} & = &
\int\limits_0^1 \frac{d \hat z \ \Theta(\hat z)}{1 - \hat z} \left( \frac{1}{\hat z} + \frac{{\bf q}_{T 2}^2}{q^2} \right)
\frac{3 \delta^{(4)}(q_1 - q_3 - q)}{32 \pi q_3^+ q_2^- (- q^2)}
\nonumber \\ & \times &
\frac{d \Psi_\gamma({\bf q}_{T 1}^2, {\bf q}_T^2)}{d {\bf q}_{T 3}^2 d y_3}
\frac{d \Psi_i({\bf q}_T^2, {\bf q}_{T 2}^2)}{{\bf q}_{T 4}^2 d {\bf q}_{T 4}^2},
\label{eq:HSCs}
\end{eqnarray}
where photon IF describes $Q \, (q_1) + \bar Q \, (q) \to \gamma  \, (q_3)$:
\begin{equation}
\frac{d \Psi_\gamma({\bf q}_{T 1}^2, {\bf q}_T^2)}{d {\bf q}_{T 3}^2 d y_3} =
4 \pi \alpha \, \frac{C_A}{N_c^2} \left( {\bf q}_{T 1}^2 + {\bf q}_T^2 \right),
\label{eq:IF-A}
\end{equation}
quark IF stands for $Q \, (q) + R \, (q_2) \to q \, (q_4)$:
\begin{equation}
\frac{d \Psi_q({\bf q}_T^2, {\bf q}_{T 2}^2)}{d {\bf q}_{T 4}^2} =
4 \pi \alpha_s \, \frac{C_A C_F}{N_c (N_c^2 - 1)} \left( {\bf q}_T + {\bf q}_{T 2} \right)^2,
\label{eq:IF-q}
\end{equation}
gluon IF $\Psi_g({\bf q}_T^2, {\bf q}_{T 2}^2)$ for $Q \, (q) + \bar Q \, (q_2) \to g \, (q_4)$ can be easily obtained from Eq.~(\ref{eq:IF-A}).
In the Regge limit $\hat z \to 0$, the integral in Eq.~(\ref{eq:HSCs}) generates $\ln\left( 1 / \hat z \right)$ and subtracts one iteration of rapidity evolution.
Using Eq.~(\ref{eq:IF-q}), one can note that in the case $i = q$, Eq.~(\ref{eq:HSCs}) is IR finite, contrary to $i = g$.

For numerical calculations, we used improved Kimber-Martin-Ryskin-Watt UPDFs~\cite{PRA}.
Cross sections as functions of the rapidity difference $Y$ are shown in Fig.~\ref{fig:2}.
For the subprocess~(\ref{eq:NLO2}) we set $|{\bf q}_{T 4}| > 1$ GeV to eliminate IR divergences.
In both cases, subtraction terms~(\ref{eq:HSCs}) without the second factor in Eq.~(\ref{eq:Theta}) oversubtracts the corresponding NLO${}^\star$ terms at low $Y$; the latter is cured by taking into account the second factor in Eq.~(\ref{eq:Theta}), which suppresses subtraction terms at low $Y$ and doesn't affect at large $Y$, where NLO${}^\star$ terms get subtracted as it should be.

\section*{Conclusions}

We have considered the double counting subtraction procedure in the application to the single photon production at NLO${}^\star$ approximation of the HEF approach.
This procedure can be improved by introducing second $\theta$-function in Eq.~(\ref{eq:Theta}), which eliminates the oversubtraction at low $Y$.

\section*{Acknowledgments}
The work is supported by the BASIS Foundation for the Advancement of Theoretical Physics and Mathematics, grant \#24-1-1-16-5.


\begin{thebibliography}{1}

\bibitem{HEF}
L.~V.~Gribov, E.~M.~Levin and M.~G.~Ryskin, Phys. Rept. \textbf{100} (1983), 1-150;
S.~Catani, M.~Ciafaloni and F.~Hautmann, Nucl. Phys. B \textbf{366} (1991), 135-188;
J.~C.~Collins and R.~K.~Ellis, Nucl. Phys. B \textbf{360} (1991), 3-30;

\bibitem{HEF2}
J.~P.~Lansberg, M.~Nefedov and M.~A.~Ozcelik, JHEP \textbf{05} (2022), 083;
J.~P.~Lansberg, M.~Nefedov and M.~A.~Ozcelik, Eur. Phys. J. C \textbf{84} (2024) no.4, 351;
C.~A.~Flett, J.~P.~Lansberg, S.~Nabeebaccus, M.~Nefedov, P.~Sznajder and J.~Wagner, Phys. Lett. B \textbf{859} (2024), 139117;
A.~A.~Chernyshev, M.~A.~Nefedov and V.~A.~Saleev, [arXiv:2506.10458 [hep-ph]];

\bibitem{HN25}
A.~van Hameren and M.~Nefedov, JHEP \textbf{02} (2025), 160;

\bibitem{DC2}
M.~A.~Nefedov, JHEP \textbf{08} (2020), 055;

\bibitem{DC}
J.~Bartels, A.~Sabio Vera and F.~Schwennsen, JHEP \textbf{11} (2006), 051;
M.~Hentschinski and A.~Sabio Vera, Phys. Rev. D \textbf{85} (2012), 056006;

\bibitem{BFKL}
E.~A.~Kuraev, L.~N.~Lipatov and V.~S.~Fadin, Sov. Phys. JETP \textbf{45} (1977), 199-204;
I.~I.~Balitsky and L.~N.~Lipatov, Sov. J. Nucl. Phys. \textbf{28} (1978), 822-829;

\bibitem{PRA2}
V.~A.~Saleev, Phys. Rev. D \textbf{78} (2008), 034033;
M.~A.~Nefedov and V.~A.~Saleev, Phys. Rev. D \textbf{92} (2015) no.9, 094033;
A.~Karpishkov and V.~Saleev, Phys. Rev. D \textbf{106} (2022) no.5, 5;
A.~Chernyshev and V.~Saleev, Phys. Rev. D \textbf{110} (2024) no.11, 11;

\bibitem{PRA}
M.~A.~Nefedov, V.~A.~Saleev and A.~V.~Shipilova, Phys. Rev. D \textbf{87} (2013) no.9, 094030;
M.~A.~Nefedov and V.~A.~Saleev, Phys. Rev. D \textbf{102} (2020), 114018.

\end{thebibliography}
\end{document}